\documentclass[english]{article}
\usepackage[T1]{fontenc}
\usepackage[latin9]{inputenc}
\usepackage{geometry}
\usepackage{color}
\usepackage{amsmath}
\usepackage{amssymb}
\usepackage{graphicx}
\usepackage{esint}
\usepackage[numbers]{natbib}

\makeatletter
\newcommand{\lyxaddress}[1]{
\par {\raggedright #1
\vspace{1.4em}
\noindent\par}
}

\@ifundefined{date}{}{\date{}}
\@ifundefined{showcaptionsetup}{}{%
 \PassOptionsToPackage{caption=false}{subfig}}
\usepackage{subfig}
\makeatother

\usepackage{babel}
\begin{document}

\title{Calculation of the current response in a nanojunction for an arbitrary
time-dependent bias: application to the molecular wire}

\author{Michael Ridley,$^{1,2}$ Angus MacKinnon,$^{2}$ and Lev Kantorovich$^{1}$}

\maketitle

\lyxaddress{$^{1}$Department of Physics, King's College London, Strand, London,
WC2R 2LS, United Kingdom\\$^{2}$Department of Physics, The Blackett
Laboratory, Imperial College London, South Kensington Campus, London
SW7 2AZ, United Kingdom}
\begin{abstract}
Recently {[}Phys. Rev. B \textbf{\textcolor{black}{91}}, 125433 (2015){]}
we derived a general formula for the time-dependent quantum electron
current through a molecular junction subject to an arbitrary time-dependent
bias within the Wide Band Limit Approximation (WBLA) and assuming
a single particle Hamiltonian. Here we present an efficient numerical
scheme for calculating the current and particle number. Using the
Pad\'{e} expansion of the Fermi function, it is shown that all frequency
integrals occurring in the general formula for the current can be
removed analytically. Furthermore, when the bias in the reservoirs
is assumed to be sinusoidal it is possible to manipulate the general
formula into a form containing only summations over special functions.
To illustrate the method, we consider electron transport through a
one-dimensional molecular wire coupled to two leads subject to out-of-phase
biases. We also investigate finite size effects in the current response
and particle number that results from the switch-on of such a bias.
\end{abstract}

\section{Introduction}

The Landauer-B\"{u}ttiker (LB) formalism has for some time provided a
reliable method for modelling coherent transport through molecular
junctions \citep{LANDAUER1970,BUTTIKER1992}. In the original formulation,
this method enabled computation of the current in the leads coupled
to a molecular region C, when a bias in at least one of the leads
causes a constant shift in the Fermi energies of the leads. Work done
with the LB formalism has mainly focused on the calculation of steady-state
properties, corresponding to a time regime that consigns the bias
switch-on time to the distant past \citep{diventra}. The LB expressions
for the current and quantum noise may be derived from the Nonequilibrium
Green's Function (NEGF) formalism, which is a more general alternative
to the S-Matrix approach \citep{RVLS2013,JAUHO1994}. This requires
the solution of the Kadanoff-Baym equations for the lesser Green's
function of the molecular region \citep{RVLS2013}, from which the
electric current in the leads and electron number in the molecule
may be extracted. When the bias is time-independent, all GFs are functions
of the time difference, and hence it is possible to work purely in
the frequency domain when solving for the components that enter the
expression for the current.

However, recent studies have opened up the possibility of calculating
the time-dependent (TD) current in a multilevel nanojunction using
a closed integral formula; expressions that include the transient
effects of the switch-on (and hence the system preparation) have been
derived for the cases of a constant \citep{RVLS2013,RVL2013,RVL2014}
and time-dependent \citep{Ridley2015} bias. Common to both approaches
is the assumption of the Wide-Band Limit Approximation (WBLA), which
neglects the detailed electronic structure of the leads and renders
the Kadanoff-Baym equations for the lesser GF analytically tractable.
Computationally, the main cost of the method we presented in \citep{Ridley2015}
lies with the evaluation of several integrals in the $\left(\omega,t\right)$
plane. The purpose of this paper is to present an efficient computational
scheme for evaluating these integrals with no loss of accuracy making
consideration of complex multi-level and multi-terminal systems within
the WBLA numerically possible. 

The structure of this paper is as follows: in Section 2, we introduce
the model of the generic switch-on problem and the basic assumptions
involved in the WBLA. In addition we prove a continuity equation for
the current formula derived in \citep{Ridley2015}. In Section 3,
the formula for the current is re-expressed using the Pad\'{e} expansion
of the Fermi function to remove all frequency integrals, forming the
basis for our numerical implementation. Section 4 presents numerical
results for the current through a molecular wire, modelled with a
simple tight-binding Hamiltonian. In this implementation, expressions
are worked out for a sinusoidal bias whereby Bessel function expansions
are used to remove all $t-$integrals in the TD current. This enables
simulations of the transport resulting from any combination of sinusoidal
biases in the leads, including those which contain a symmetry-breaking
phase. Furthermore it enables us to identify some novel finite size
effects on the dynamics and to study the conservation of charge as
the transient current dies out.

\section{The Model \label{sec:The-Model}}

\subsection{Hamiltonian and GF Components}

The NEGF formalism is a method for calculating the evolution in time
of ensemble averages, following an event at a particular time $t_{0}$
which breaks the physical symmetry of the system between times $t=\pm\infty$.
In quantum transport one is primarily concerned with calculations
of the time-dependent current following the switch-on of a bias in
one of the leads. Here we model this switch-on by adding a spatially
constant time-dependent shift to the lead energy levels. 

We will in general be concerned with the following Hamiltonian:

\begin{gather}
\hat{H}\left(t\right)=\underset{k\alpha}{\sum}\varepsilon_{k\alpha}\left(t\right)\hat{d}_{k\alpha}^{\dagger}\hat{d}_{k\alpha}+\underset{mn}{\sum}V_{mn}\hat{d}_{m}^{\dagger}\hat{d}_{n}+\underset{m,k\alpha}{\sum}\left[V_{m,k\alpha}\hat{d}_{m}^{\dagger}\hat{d}_{k\alpha}+V_{k\alpha,m}\hat{d}_{k\alpha}^{\dagger}\hat{d}_{m}\right]\label{eq:Hamiltonian}
\end{gather}
Here, $\hat{d}_{k\alpha}$, $\hat{d}_{m}$ and $\hat{d}_{k\alpha}^{\dagger}$,
$\hat{d}_{m}^{\dagger}$ are annihilation and creation operators of
leads and central system electronic states, where for simplicity spin
degrees of freedom are neglected. We collect elements of this Hamiltonian
into a matrix consisting of `blocks' corresponding to each of the
physical subsystems it describes. The first term is a Hamiltonian
of the lead states $k$ belonging to each lead $\alpha$. The second
term is the Hamiltonian of the molecule coupled to the lead; it contains
inter-molecular hopping matrix elements $V_{mn}$ between the $m$
and $n$ molecular orbitals, defining a molecular Hamiltonian matrix
`block' $\mathbf{h}_{CC}$ with elements $V_{mn}$. The third term
describes the coupling of the molecule to the leads, and defines the
$\alpha-C$ `block' of the Hamiltonian matrix, $\mathbf{h}_{\alpha C}$
with elements $V_{k\alpha,m}$. The switch-on problem is specified
by assigning values to the matrix elements of $\hat{H}\left(t\right)$
at times before and after $t_{0}$. 

We assume that, prior to $t_{0}$, the system has equilibrated with
lead-molecule couplings present in the Hamiltonian $\hat{H}\left(t<t_{0}\right)\equiv\hat{H}_{0}$,
so that $\hat{H}_{0}$ is as in Eq. (\ref{eq:Hamiltonian}) with time-independent
lead state energies $\varepsilon_{k\alpha}\left(t<t_{0}\right)=\varepsilon_{k\alpha}$.
The system is therefore described by an equilibrium density operator
$\widehat{\rho}_{0}=Z^{-1}e^{-\beta\left(\widehat{H}_{0}-\mu\widehat{N}\right)}$,
where $Z$ is the partition function, $\hat{N}$ the number operator,
$\beta=1/k_{B}T$ the inverse temperature and $\mu$ the chemical
potential. At times following the switch-on, the lead energies acquire
a time-dependent shift, $\varepsilon_{k\alpha}\left(t>t_{0}\right)=\varepsilon_{k\alpha}+V_{\alpha}\left(t\right)$,
where the function $V_{\alpha}\left(t\right)$ is arbitrary. This
is an example of a \textit{partition-free }approach \citep{StefALm2004,RVL2013,Ridley2015}
to the switch-on problem, as the system Hamiltonian is not divided
into decoupled subregions prior to $t_{0}$. \textit{Partitioned}
approaches \citep{JAUHO1994} add the lead-molecule coupling at the
switch-on time and therefore cannot be expected to give a physically
meaningful transient.

The problem of calculating the current in response to this kind of
switch-on is mapped to the solution of a set of coupled integro-differential
equations for Green's Functions (GFs) defined on a complex time contour
$\gamma$, which we refer to as the Konstantinov-Perel' contour \citep{konstantinov-perel-JETP-1961}.
One then solves for various components of the two-time GF of the central
region $CC$ `block', whose rigorous definition may be found elsewhere
\citep{RVLS2013,Ridley2015}. Here we simply note that the retarded
GF satisfies the following equation of motion: 

\begin{eqnarray}
\left(i\frac{d}{dt_{1}}-\mathbf{h}_{CC}\right)\mathbf{G}_{CC}^{r}\left(t_{1},t_{2}\right) & = & \mathbf{1}_{CC}\delta\left(t_{1}-t_{2}\right)+\underset{t_{0}}{\overset{t}{\int}}d\bar{t}\,\mathbf{\Sigma}_{CC}^{r}\left(t_{1},\bar{t}\right)\mathbf{G}_{CC}^{r}\left(\bar{t},t_{2}\right)\label{eq:EoM-retarded}
\end{eqnarray}
On the right-hand side of this expression, one encounters the retarded
component of the embedding self-energy,

\begin{equation}
\left[\mathbf{\Sigma}_{CC}^{r}\left(t_{1},t_{2}\right)\right]_{mn}=\sum_{\alpha}e^{-i\psi_{\alpha}\left(t_{1},t_{2}\right)}\int\frac{d\omega}{2\pi}e^{i\omega\left(t_{1}-t_{2}\right)}\left[\Lambda_{\alpha,mn}\left(\omega\right)-\frac{i}{2}\Gamma_{\alpha,mn}\left(\omega\right)\right]\label{eq:self-energy}
\end{equation}
We have in this expression introduced the level width matrix 
\begin{equation}
\mathbf{\Gamma}_{\alpha,mn}\left(\omega\right)=2\pi\sum_{k}T_{m,k\alpha}T_{k\alpha,n}\delta\left(\omega-\epsilon_{k\alpha}\right)
\end{equation}
which forms a Hilbert transform pair with $\Lambda_{\alpha,mn}\left(\omega\right)$.
In addition we have defined the phase factor: 
\begin{equation}
\psi_{\alpha}\left(t_{1},t_{2}\right)=\underset{t_{1}}{\overset{t_{2}}{\int}}d\bar{t}\, V_{\alpha}\left(\bar{t}\right)
\end{equation}
In the WBLA, one assumes that $\mathbf{\Gamma}_{\alpha,mn}\left(\omega\right)$
is replaced with a frequency-independent value $\varGamma_{\alpha}=\mathbf{\Gamma}_{\alpha,mn}\left(\varepsilon_{\alpha}^{F}\right)$,
where $\varepsilon_{\alpha}^{F}$ is the equilibrium Fermi energy
of lead $\alpha$ \citep{JAUHO1994}. This has the effect of mapping
Eq. (\ref{eq:EoM-retarded}) onto an expression which is analytically
tractable, resulting in the following exact expression:

\begin{equation}
\mathbf{G}_{CC}^{r}\left(t_{1},t_{2}\right)=\int\frac{d\omega}{2\pi}e^{i\omega\left(t_{1}-t_{2}\right)}\left(\omega\mathbf{1}_{CC}-\mathbf{h}_{CC}^{eff}\right)^{-1}\equiv\int\frac{d\omega}{2\pi}e^{i\omega\left(t_{1}-t_{2}\right)}\mathbf{G}_{CC}^{r}\left(\omega\right)\label{eq:retardedGF}
\end{equation}
In (\ref{eq:retardedGF}), $\mathbf{h}_{CC}^{eff}\equiv\mathbf{h}_{CC}-\frac{i}{2}\underset{\alpha}{\sum}\mathbf{\Gamma}_{\alpha}$
is an effective Hamiltonian of the central region, whose eigenvalues
correspond to unstable eigenmodes of the molecular structure, which
have acquired a finite lifetime due to the presence of the leads.
All other components of the GF can also be explicitly calculated in
the time domain \citep{Ridley2015}. In particular, the greater and
lesser GFs can be expressed as 

\begin{equation}
\mathbf{G}_{CC}^{\gtrless}\left(t_{1},t_{2}\right)=\mp i\int\frac{d\omega}{2\pi}f\left(\mp\left(\omega-\mu\right)\right)\underset{\alpha}{\sum}\mathbf{S}_{\alpha}\left(t_{1},t_{0};\omega\right)\mathbf{\Gamma}_{\alpha}\mathbf{S}_{\alpha}^{\dagger}\left(t_{2},t_{0};\omega\right)\label{eq:greater/lesserGF}
\end{equation}
where we make the definition 
\begin{equation}
\mathbf{S}_{\alpha}\left(t,t_{0};\omega\right)\equiv e^{-i\mathbf{h}_{CC}^{eff}\left(t-t_{0}\right)}\left[\mathbf{G}_{CC}^{r}\left(\omega\right)-i\int_{t_{0}}^{t}d\bar{t}e^{-i\left(\omega\mathbf{1}-\mathbf{h}_{CC}^{eff}\right)\left(\bar{t}-t_{0}\right)}e^{-i\psi_{\alpha}\left(\bar{t},t_{0}\right)}\right]
\end{equation}
and note that all information concerning the time-dependent bias $V_{\alpha}\left(t\right)$
of lead $\alpha$ is to be found in the phase factor $\psi_{\alpha}\left(\bar{t},t_{0}\right)$.

\subsection{Current and Continuity Equation}

In \citep{Ridley2015}, results were presented for the transport through
a quantum dot using direct numerical integration to evaluate the current
at each time $t>t_{0}$. The current in lead $\alpha$ is defined
as the time derivative of the average particle number in that lead,
$I_{\alpha}\left(t\right)\equiv q\left\langle \frac{d\hat{N}_{\alpha}\left(t\right)}{dt}\right\rangle $,
and may be expressed in the following rather compact form (taking
electron charge $q=-1$ and a spin degeneracy of 2):

\begin{equation}
I_{\alpha}\left(t\right)=\frac{1}{\pi}\int d\omega f\left(\omega-\mu\right)\,\mbox{Tr}_{C}\left[2\mbox{Re}\left[ie^{i\omega\left(t-t_{0}\right)}e^{i\psi_{\alpha}\left(t,t_{0}\right)}\mathbf{S}_{\alpha}\left(t,t_{0};\omega\right)\mathbf{\Gamma}_{\alpha}\right]-\mathbf{\Gamma}_{\alpha}\underset{\beta}{\sum}\mathbf{S}_{\beta}\left(t,t_{0};\omega\right)\mathbf{\Gamma}_{\beta}\mathbf{S}_{\beta}^{\dagger}\left(t,t_{0};\omega\right)\right]\label{eq: CURRENT}
\end{equation}
The molecule which has been sandwiched between macroscopic leads is
fundamentally an open quantum system, so the total charge in the molecular
region is not necessarily conserved even though the total amount of
charge in the molecular region plus that of all the leads is fixed
for all times. This is a condition automatically satisfied by the
NEGF formalism \citep{RVLS2013}. The condition for local charge conservation
is that the sum of all currents measured in the leads of a multiterminal
device is equal to zero:

\begin{equation}
\underset{\alpha}{\sum}I_{\alpha}\left(t\right)=0\label{eq:conservation}
\end{equation}
This condition states that the rate of charge entering the molecular
region is balanced by the rate of charge leaving it. One should now
recall that there is no direct coupling between leads, so that all
charge transfer between leads is mediated by the central region, and
also that the definition of the current in lead $\alpha$ was given
as $I_{\alpha}\left(t\right)\equiv q\left\langle \frac{d\hat{N}_{\alpha}\left(t\right)}{dt}\right\rangle $,
i.e. it was given as the rate of \textit{increase} of charge in this
lead. Eq. (\ref{eq:conservation}) is simply a statement of the condition
that if charge increases in one lead connected to the molecule, then
this increase is compensated exactly by a net decrease of charge from
the other leads making up the junction. The total amount of charge
in the molecular region at any given time is directly related to the
lesser GF:

\begin{eqnarray}
N_{C}\left(t\right) & = & -2i\mbox{Tr}_{C}\left[\mathbf{G}_{CC}^{<}\left(t,t\right)\right]=\frac{1}{\pi}\underset{\beta}{\sum}\int d\omega f\left(\omega-\mu\right)\mbox{Tr}_{C}\left[\mathbf{S}_{\beta}\left(t,t_{0};\omega\right)\mathbf{\Gamma}_{\beta}\mathbf{S}_{\beta}^{\dagger}\left(t,t_{0};\omega\right)\right]\label{eq:numbermol}
\end{eqnarray}
The time derivative of the charge number $\frac{dN_{C}\left(t\right)}{dt}$
in the central region has been referred to as the \textit{displacement
current} in the literature \citep{RVLS2013}. To make precise the
relation between this quantity and the currents measured in the leads,
we take the time-derivatives of the matrix $\mathbf{S}_{\beta}\equiv\mathbf{S}_{\beta}\left(t,t_{0};\omega\right)$
and its complex conjugate, e.g.

\begin{eqnarray}
\frac{d\mathbf{S}_{\beta}}{dt} & = & -i\mathbf{h}_{CC}^{eff}\mathbf{S}_{\beta}-ie^{-i\omega\left(t-t_{0}\right)}e^{-i\psi_{\beta}\left(t,t_{0}\right)}
\end{eqnarray}
This expression, along with its complex conjugate, is substituted
into the derivative of $N_{C}$ in Eq. (\ref{eq:numbermol}) to give

\begin{eqnarray*}
\frac{d\mathbf{N}_{C}\left(t\right)}{dt} & = & \frac{1}{\pi}\underset{\beta}{\sum}\int d\omega f\left(\omega-\mu\right)\mbox{Tr}_{C}\left\{ 2\mbox{Re}\left[ie^{i\omega\left(t-t_{0}\right)}e^{i\psi_{\beta}\left(t,t_{0}\right)}\mathbf{S}_{\beta}\mathbf{\Gamma}_{\beta}\right]-i\mathbf{S}_{\beta}\mathbf{\Gamma}_{\beta}\mathbf{S}_{\beta}^{\dagger}\left(\mathbf{h}_{CC}^{eff}-\mathbf{h}_{CC}^{eff\dagger}\right)\right\} \\
 & = & \frac{1}{\pi}\underset{\alpha}{\sum}\int d\omega f\left(\omega-\mu\right)\mbox{Tr}_{C}\left\{ 2\mbox{Re}\left[ie^{i\omega\left(t-t_{0}\right)}e^{i\psi_{\alpha}\left(t,t_{0}\right)}\mathbf{S}_{\beta}\mathbf{\Gamma_{\beta}}\right]-\sum_{\alpha}\mathbf{\Gamma}_{\alpha}\mathbf{S}_{\beta}\mathbf{\Gamma}_{\beta}\mathbf{S}_{\beta}^{\dagger}\right\} 
\end{eqnarray*}
where we use the fact that $\mathbf{h}_{CC}^{eff}-\mathbf{h}_{CC}^{eff\dagger}=-i\underset{\alpha}{\sum}\mathbf{\Gamma}_{\alpha}\equiv-i\mathbf{\Gamma}$.
If we then sum over lead indices in the expression (\ref{eq: CURRENT}),
the following theorem is established:

\begin{equation}
\frac{d\mathbf{N}_{C}\left(t\right)}{dt}=\underset{\alpha}{\sum}I_{\alpha}\left(t\right)\label{eq: ConservationLaw}
\end{equation}
This is a statement of global charge conservation in the nanojunction.
When the bias in every lead is constant, (\ref{eq:numbermol}) reduces
to the formula for the particle number given in \citep{RVL2013}.
In the long time limit the number of charges on the molecule is constant
in time, i.e. an ideal stationary state is reached; hence, trivially
the condition for local charge conservation (\ref{eq:conservation})
is satisfied in this case. Although the condition of global charge
conservation (\ref{eq: ConservationLaw}) is never violated, physical
situations exist in which the condition (\ref{eq:conservation}) for
local charge conservation in the molecular (central) region of a nanojunction
is violated, i.e. $dN_{C}(t)/dt\neq0$.

\section{Pade Expansion of the Time-Dependent Current\label{sec:SEC3}}

\subsection{Exact Formula for the Current: Hurwitz-Lerch Functions}

In previous work \citep{Ridley2015}, numerical integration was used
to compute the current formula (\ref{eq: CURRENT}) at different times
for a single level system. We now seek to extend this work to a numerical
method that facilitates time-dependent transport calculations in systems
of real experimental interest, for instance carbon nanotube transistors
\citep{Li2004}. To proceed with a numerical implementation of Eqs.
(\ref{eq: CURRENT}) and (\ref{eq:numbermol}), one can introduce
the right and left eigenproblems for the renormalized Hamiltonian
matrix $\mathbf{h}_{CC}^{eff}$ \citep{RVLS2014}:

\begin{eqnarray}
\mathbf{h}_{CC}^{eff}\left|\varphi_{j}^{R}\right\rangle  & = & \bar{\varepsilon}_{j}\left|\varphi_{j}^{R}\right\rangle \,\,\mbox{and}\,\,\left\langle \varphi_{j}^{L}\right|\mathbf{h}_{CC}^{eff}=\bar{\varepsilon}_{j}\left\langle \varphi_{j}^{L}\right|
\end{eqnarray}
The eigenenergies $\bar{\varepsilon}_{j}$ contain an imaginary part
that is strictly negative (as $\mathbf{\Gamma}$ is positive-definite),
and the same value of $\bar{\varepsilon}_{j}$ corresponds to each
of the left and right eigenvectors. Using the idempotency property
\begin{equation}
\underset{j}{\sum}\frac{\left|\varphi_{j}^{R}\right\rangle \left\langle \varphi_{j}^{L}\right|}{\left\langle \varphi_{j}^{L}\mid\varphi_{j}^{R}\right\rangle }=\mathbf{I}=\underset{j}{\sum}\frac{\left|\varphi_{j}^{L}\right\rangle \left\langle \varphi_{j}^{R}\right|}{\left\langle \varphi_{j}^{R}\mid\varphi_{j}^{L}\right\rangle }\;,
\end{equation}
the formula (\ref{eq: CURRENT}) can be put into the form:

\begin{eqnarray}
I_{\alpha}\left(t\right) & = & \frac{1}{\pi}\underset{j}{\sum}\left\{ \int dx\, f\left(x\right)2\mbox{Re}\left[\frac{\left\langle \varphi_{j}^{L}\right|\mathbf{\Gamma}_{\alpha}\left|\varphi_{j}^{R}\right\rangle }{\left\langle \varphi_{j}^{L}\right.\left|\varphi_{j}^{R}\right\rangle }\left(i\frac{e^{i\psi_{\alpha}\left(t,t_{0}\right)}e^{i\left(x+\mu-\bar{\varepsilon}_{j}\right)\left(t-t_{0}\right)}}{x+\mu-\bar{\varepsilon}_{j}}+\underset{t_{0}}{\overset{t}{\int}}d\bar{t}\, e^{i\psi_{\alpha}\left(t,\bar{t}\right)}e^{i\left(x+\mu-\bar{\varepsilon}_{j}\right)\left(t-\bar{t}\right)}\right)\right]\right.\label{eq: CURRENT2}\\
 &  & -\underset{\beta,k}{\sum}\frac{\left\langle \varphi_{k}^{R}\right|\mathbf{\Gamma}_{\alpha}\left|\varphi_{j}^{R}\right\rangle \left\langle \varphi_{j}^{L}\right|\mathbf{\Gamma}_{\beta}\left|\varphi_{k}^{L}\right\rangle }{\left\langle \varphi_{j}^{L}\right.\left|\varphi_{j}^{R}\right\rangle \left\langle \varphi_{k}^{R}\right.\left|\varphi_{k}^{L}\right\rangle }e^{-i\left(\bar{\varepsilon}_{j}-\bar{\varepsilon}_{k}^{*}\right)\left(t-t_{0}\right)}\left[\underset{t_{0}}{\overset{t}{\int}}d\bar{t}\underset{t_{0}}{\overset{t}{\int}}d\bar{t}'\, e^{-i\psi_{\beta}\left(\bar{t},\bar{t}'\right)}e^{-i\left(x+\mu-\bar{\varepsilon}_{j}\right)\left(\bar{t}-t_{0}\right)}e^{i\left(x+\mu-\bar{\varepsilon}_{k}^{*}\right)\left(\bar{t}'-t_{0}\right)}\right.\nonumber \\
 &  & \left.\left.+\frac{1}{\left(x+\mu-\bar{\varepsilon}_{j}\right)\left(x+\mu-\bar{\varepsilon}_{k}^{*}\right)}+\underset{t_{0}}{\overset{t}{\int}}d\bar{t}\left[i\frac{e^{i\psi_{\beta}\left(\bar{t},t_{0}\right)}e^{i\left(x+\mu-\bar{\varepsilon}_{k}^{*}\right)\left(\bar{t}-t_{0}\right)}}{x+\mu-\bar{\varepsilon}_{j}}+c.c_{j\leftrightarrow k}\right]\right]\right\} \nonumber 
\end{eqnarray}
where $f\left(x\right)$ is the Fermi function and $c.c_{j\leftrightarrow k}$
denotes the complex conjugate of the preceding term with indices $j$
and $k$ exchanged. We therefore reduce $I_{\alpha}\left(t\right)$
to a sum of integrals over scalar functions in the $\left(x,t\right)$
plane. One integral whose structure is repeated throughout (\ref{eq: CURRENT2})
can be performed analytically, using a contour integral over a semi-circular
contour in the upper-half of the complex plane:

\begin{equation}
\int dx\frac{f\left(x\right)e^{i\left(x-\left(\bar{\varepsilon}_{j}-\mu\right)\right)\left(t-t_{0}\right)}}{x-\left(\bar{\varepsilon}_{j}-\mu\right)}=-e^{-\frac{\pi}{\beta}\left(t-t_{0}\right)}e^{-i\left(\bar{\varepsilon}_{j}-\mu\right)\left(t-t_{0}\right)}\Phi\left(e^{-\frac{2\pi}{\beta}\left(t-t_{0}\right)},1,\frac{1}{2}-\frac{\beta}{2i\pi}\left(\bar{\varepsilon}_{j}-\mu\right)\right)\label{eq: INT}
\end{equation}
Here we have introduced the so-called \textit{Hurwitz-Lerch Transcendent}\textcolor{black}{{}
$\Phi$ \citep{lerch1887note}:}

\begin{equation}
\Phi\left(z,s,a\right)\equiv\underset{n=0}{\overset{\infty}{\sum}}\frac{z^{n}}{\left(n+a\right)^{s}}\label{eq:HLT}
\end{equation}
The formula (\ref{eq: INT}) enables us to replace several frequency
integrals with a fast-converging series expansi\textcolor{black}{on.
We note in passing that, if one follows the steps in the formal integration
of the lesser GF $\mathbf{G}_{CC}^{<}\left(t_{1},t_{2}\right)$ which
enabled the derivation of Eq. (\ref{eq: CURRENT}) in Ref. \citep{Ridley2015},
one uses a well-known \citep{RVLS2013} transformation of Matsubara
summations into frequency integrals when evaluating a set of convolution
integrals taken along the vertical branch of the Konstantinov-Perel'
contour. However, these steps can be omitted altogether, and a direct
route can be taken to an expression for $I_{\alpha}\left(t\right)$
in terms of $\Phi$ which retains the Matsubara summation at all stages. }

\subsection{Removal of Frequency Integrals: Pad\'{e} Expansion}

The Pad\'{e} approximation to the Fermi function is a series expansion
whose terms possess a simple pole structure \citep{Ozaki2007,Hu2010,Hu2011}:

\begin{equation}
f\left(x\right)=\frac{1}{e^{\beta x}+1}=\frac{1}{2}-\underset{N\rightarrow\infty}{\lim}\underset{l=1}{\overset{N}{\sum}}\eta_{l}\left(\frac{1}{\beta x+i\zeta_{l}}+\frac{1}{\beta x-i\zeta_{l}}\right)\label{eq:PADE}
\end{equation}
This is identical in structure to the Matsubara expansion, which has
parameter values $\eta_{l}=1$ and $\zeta_{l}=\pi\left(2l-1\right)$,
but it can be shown to converge much more quickly than the Matsubara
expansion as $N$ increases \citep{Hu2010,Hu2011}. Unlike the Matsubara
expansion, in which the poles are spaced at constant intervals of
$\frac{2\pi}{\beta}$ along the imaginary axis, the poles $i\zeta_{l}/\beta$
in the Pad\'{e} expansion are spaced unevenly along the imaginary axis,
and the prefactors $\eta_{l}$ are all real and positive-valued. In
practice, the expansion $f_{N}(x)$ truncated at some finite value
of $N$ is used, with $N$ chosen such that for values of $L\gg\varepsilon$,
where $\varepsilon$ denotes the typical energy scale of the problem,
the deviation $\delta f_{N}\left(L\right)\equiv\left|f\left(x\right)-f_{N}\left(x\right)\right|_{\beta x=L}<10^{-p}$,
where $p$ can be chosen to give arbitrary accuracy \citep{Hu2010,Hu2011,zhang(b)2013}.
In practice we find good convergence for $N\sim20$ when the Pad\'{e}
parameters are used.

The expansion (\ref{eq:PADE}) is then substituted into (\ref{eq: CURRENT2}),
resulting in $x$-integrals that can be evaluated analytically with
the residue theorem, for example:

\begin{equation}
\int dxf\left(x\right)e^{ix\left(t-\bar{t}\right)}=\pi\left[\delta\left(t-\bar{t}\right)-\frac{2i}{\beta}\underset{N\rightarrow\infty}{\lim}\underset{l=1}{\overset{N}{\sum}}\eta_{l}e^{-\frac{\zeta_{l}}{\beta}\left(t-\bar{t}\right)}\right]\label{eq:INT2}
\end{equation}
Using this method and the identity (\ref{eq: INT}) one finally obtains
an expression for the current in lead $\alpha$ that is asymptotically
exact (as $N\rightarrow\infty$), and which has all frequency integrals
eliminated:

\begin{gather}
I_{\alpha}\left(t\right)\simeq\frac{1}{\pi}\underset{j}{\sum}\left[2\mbox{Re}\left[\frac{\left\langle \varphi_{j}^{L}\right|\mathbf{\Gamma}_{\alpha}\left|\varphi_{j}^{R}\right\rangle }{\left\langle \varphi_{j}^{L}\mid\varphi_{j}^{R}\right\rangle }\left(-ie^{i\psi_{\alpha}\left(t,t_{0}\right)}e^{-i\left(\bar{\varepsilon}_{j}-\mu-i\frac{\pi}{\beta}\right)\left(t-t_{0}\right)}\Phi\left(e^{-\frac{2\pi}{\beta}\left(t-t_{0}\right)},1,\frac{1}{2}-\frac{\beta}{2i\pi}\left(\bar{\varepsilon}_{j}-\mu\right)\right)\right.\right.\right.\label{eq:CURRENTFINAL}\\
\left.\left.-\frac{2i\pi}{\beta}\underset{l=1}{\overset{N}{\sum}}\eta_{l}\underset{t_{0}}{\overset{t}{\int}}d\bar{t}e^{-i\left(\bar{\varepsilon}_{j}-\mu-i\frac{\zeta_{l}}{\beta}\right)\left(t-\bar{t}\right)}e^{i\psi_{\alpha}\left(t,\bar{t}\right)}\right)\right]\nonumber \\
-\underset{\beta,k}{\sum}\frac{\left\langle \varphi_{k}^{R}\right|\mathbf{\Gamma}_{\alpha}\left|\varphi_{j}^{R}\right\rangle \left\langle \varphi_{j}^{L}\right|\mathbf{\Gamma}_{\beta}\left|\varphi_{k}^{L}\right\rangle }{\left\langle \varphi_{j}^{L}\mid\varphi_{j}^{R}\right\rangle \left\langle \varphi_{k}^{R}\mid\varphi_{k}^{L}\right\rangle }e^{-i\left(\bar{\varepsilon}_{j}-\bar{\varepsilon}_{k}^{*}\right)\left(t-t_{0}\right)}\left[\frac{\Psi\left(\frac{1}{2}-\frac{\beta}{2i\pi}\left(\bar{\varepsilon}_{j}-\mu\right)\right)-\Psi\left(\frac{1}{2}+\frac{\beta}{2i\pi}\left(\bar{\varepsilon}_{k}^{*}-\mu\right)\right)}{\bar{\varepsilon}_{j}-\bar{\varepsilon}_{k}^{*}}\right.\nonumber \\
\underset{t_{0}}{\overset{t}{\int}}d\bar{t}\left[-ie^{-i\left(\bar{\varepsilon}_{k}^{*}-\mu-i\frac{\pi}{\beta}\right)\left(\bar{t}-t_{0}\right)}e^{i\psi_{\beta}\left(\bar{t},t_{0}\right)}\Phi\left(e^{-\frac{2\pi}{\beta}\left(\bar{t}-t_{0}\right)},1,\frac{1}{2}-\frac{\beta}{2i\pi}\left(\bar{\varepsilon}_{j}-\mu\right)\right)+c.c_{j\leftrightarrow k}\right]\nonumber \\
\left.-\frac{2\pi i}{\beta}\underset{l=1}{\overset{N}{\sum}}\eta_{l}\underset{t_{0}}{\overset{t}{\int}}d\bar{t}\underset{t_{0}}{\overset{t}{\int}}d\bar{t}'e^{i\left(\bar{\varepsilon}_{j}-\mu\right)\left(\bar{t}-t_{0}\right)}e^{-i\left(\bar{\varepsilon}_{k}^{*}-\mu\right)\left(\bar{t}'-t_{0}\right)}e^{-i\psi_{\beta}\left(\bar{t},\bar{t}'\right)}\left[-\theta\left(\bar{t}-\bar{t}'\right)e^{-\frac{\zeta_{l}}{\beta}\left(\bar{t}-\bar{t}'\right)}+\theta\left(\bar{t}'-\bar{t}\right)e^{-\frac{\zeta_{l}}{\beta}\left(\bar{t}'-\bar{t}\right)}\right]\right]\nonumber 
\end{gather}
Here we have introduced the \textit{digamma function} $\Psi$, defined
as the logarithmic derivative of the complex gamma function \citep{JAUHO1994}.
Importantly, the difference of two digamma functions appearing in
the expression for $I_{\alpha}(t)$ can be converted into a well converging
numerical series thanks to the property: 

\begin{equation}
\frac{\Psi\left(z_{1}\right)-\Psi\left(z_{2}\right)}{z_{1}-z_{2}}=\underset{n=0}{\overset{\infty}{\sum}}\frac{1}{\left(n+z_{1}\right)\left(n+z_{2}\right)}\label{eq:digam}
\end{equation}

The general result (\ref{eq:CURRENTFINAL}) forms the basis for all
subsequent numerical work involving the calculation of the current
response to an arbitrary time-dependent bias. In practice, the time
integrals may be performed numerically or removed analytically given
the assumption of a particular functional form for $V_{\alpha}\left(t\right)$.
Several of the decaying modes in the current can be directly identified
by inspection of (\ref{eq:CURRENTFINAL}), although the exact nature
of its long-time behaviour depends on the bias chosen. In particular,
if we separate out the real and imaginary parts of the complex eigenvalues,
$\bar{\varepsilon}_{j}=\lambda_{j}-i\gamma_{j}$, we see that modes
appearing in the single summation decay as $e^{-\gamma_{j}\left(t-t_{0}\right)}$,
whereas the prefactor governing the decay of modes in the double sum
is $e^{-\left(\gamma_{j}+\gamma_{k}\right)\left(t-t_{0}\right)}$.
This defines a timescale of $\tau\equiv Max\{1/\gamma_{j}\}$ as the
time taken for transient behaviour to vanish following the switch-on.

\section{Results\label{sec:Results}}

\begin{figure}
\subfloat[\label{fig:chain}]{\includegraphics[scale=0.35]{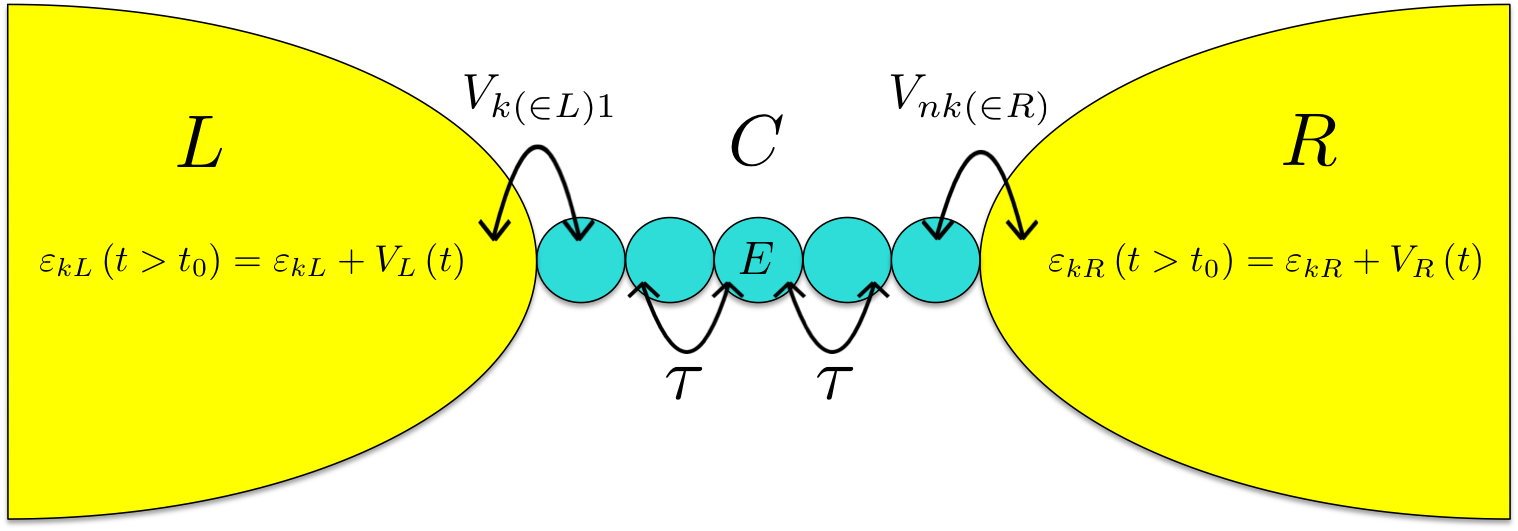}

}\subfloat[\label{fig:asym}]{\includegraphics[scale=0.25]{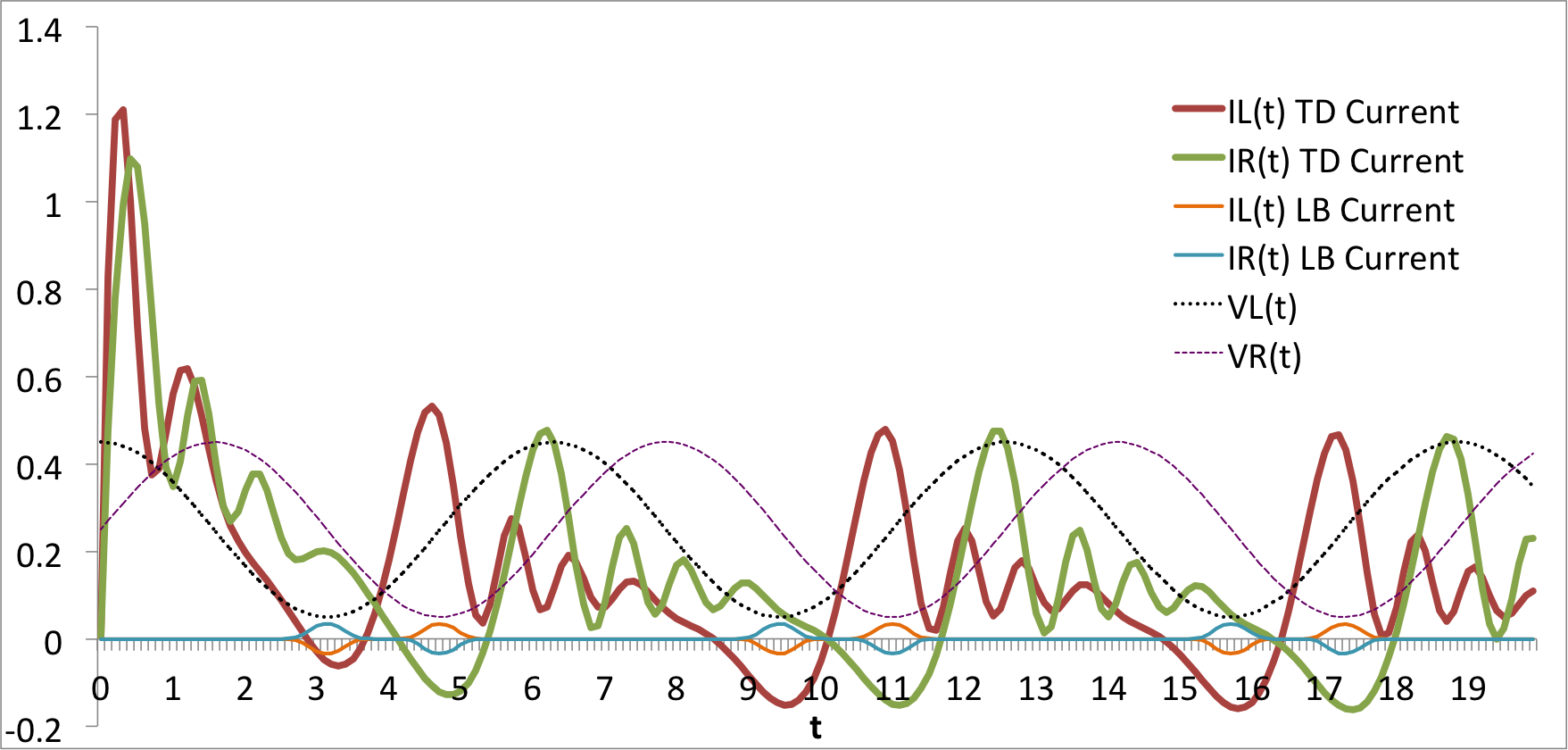}

}

\caption{(a) Schematic of the molecular chain coupled to two reservoirs, modelled
as a sequence of quantum dots with inter-dot hopping and lead-dot
hopping terms in the Hamiltonian. (b) Current through each lead of
a molecular wire junction in which the symmetry is broken due to biases
phase-shifted by $\pi/2$, i.e. $\phi_{L}=0,\,\phi_{R}=-\pi/2$. The
number of molecular sites is $5$, and other parameters chosen are
$k_{B}T=0.1,\, V_{L}=5.0=V_{R},\, A_{L}=4.0=A_{R},\,\Omega_{L}=1.0=\Omega_{R},\,\Gamma_{11}=0.5,\,\Gamma_{55}=0.5,\, E=h_{kk}=1.0,\,\tau=h_{kk+1}=h_{k+1k}=0.1$.
\label{fig:1}}
\end{figure}

We shall now apply the method which was outlined in Section \ref{sec:SEC3}
to a molecular wire coupled to two leads, illustrated schematically
in Fig. \ref{fig:1}(a). We describe this system using a tight-binding
model for the wire with nearest-neighbor hopping, as described in
\citep{MUJICA(a)1994,MUJICA(b)1994}. Specifically, we assume that
the leads are connected by a wire of 5 sites, with one state per site,
and that the interaction of the chain orbitals with those in the leads
is only via the end sites. Within the WBLA, this translates into the
condition that only the $\mathbf{\Gamma}_{11}$ and $\mathbf{\Gamma}_{55}$
elements in the level-width matrix are non-zero and are given by $\Gamma_{11}=2\pi\underset{i\in L}{\sum}\left|V_{i1}\right|^{2}\delta\left(\varepsilon_{L}^{F}-\varepsilon_{iL}\right)$
and $\Gamma_{55}=2\pi\underset{j\in R}{\sum}\left|V_{Nj}\right|^{2}\delta\left(\varepsilon_{R}^{F}-\varepsilon_{jR}\right)$,
respectively. For simplicity, we assume that for each site in the
chain there is a single energy level, so that effectively we are describing
a chain of quantum dots coupled to each other with a nearest-neighbor
hopping, similarly to the system studied in \citep{chen2013QDARRAY},
resulting in a tridiagonal molecular Hamiltonian matrix $\mathbf{h}_{CC}$,
with elements $\left[\mathbf{h}_{CC}\right]_{k,k}\equiv E$ and $\left[\mathbf{h}_{CC}\right]_{k,k+1}=\left[\mathbf{h}_{CC}\right]_{k+1,k}\equiv\tau$.
Additionally, we choose the chemical potential $\mu$ to be zero.

We now focus on the case of a sinusoidal \textcolor{black}{AC bias
turned on in each lead, with a lead-dependent constant part, amplitude,
frequency and phase shift:}

\textcolor{black}{
\begin{equation}
V_{\alpha}\left(t\right)=V_{\alpha}+A_{\alpha}\cos\left(\Omega_{\alpha}\left(t-t_{0}\right)+\phi_{\alpha}\right)\label{eq:sinusoidal}
\end{equation}
This enables us to remove analytically all the remaining integrals
in (\ref{eq:CURRENTFINAL}), by use of the identity:}

\begin{equation}
e^{i\psi_{\alpha}\left(t_{1},t_{2}\right)}=e^{iV_{\alpha}\left(t_{1}-t_{2}\right)}\underset{r,s=-\infty}{\overset{\infty}{\sum}}J_{r}\left(\frac{A_{\alpha}}{\Omega_{\alpha}}\right)J_{s}\left(\frac{A_{\alpha}}{\Omega_{\alpha}}\right)e^{i\left(r-s\right)\phi_{\alpha}}e^{ir\Omega_{\alpha}\left(t_{1}-t_{0}\right)}e^{-is\Omega_{\alpha}\left(t_{2}-t_{0}\right)}\label{eq: phase}
\end{equation}
where $J_{r}$ denotes a Bessel function of the first kind of order
$r$. When Eq. (\ref{eq: phase}) is substituted into (\ref{eq:CURRENTFINAL}),
the single and double time integrals in (\ref{eq:CURRENTFINAL}) can
be removed analytically and the problem of calculating the time-dependent
current response is reduced to a simple embedded summation over left/right
eigenvectors, the Pad\'{e} parameters and Bessel functions. For the special
case of the sinusoidal bias (\ref{eq:sinusoidal}), this summation
can be expressed entirely in terms of the special functions (\ref{eq:HLT})
and (\ref{eq:digam}). We confirm that in the case of a single quantum
dot our method reproduces exactly the $I-t$ characteristic curves
published in \citep{Ridley2015}, which were obtained by direct numerical
integration of (\ref{eq: CURRENT}). 

In this work, we apply the bias (\ref{eq:sinusoidal}) to the leads
$L$ and $R$ in the system shown in Fig. \ref{fig:1}(a). The amplitudes
$A_{\alpha}$ and constant bias shifts $V_{\alpha}$ are identical
for each lead, as is the driving frequency of the bias, $\Omega_{\alpha}$.
The physical symmetry of this transport problem is then broken by
a relative phase-shift of $\phi_{R}=-\pi/2$ of $V_{R}\left(t\right)$
with respect to $V_{L}\left(t\right)$, whose phase is fixed at $\phi_{L}=0$.
The tridiagonal central region Hamiltonian has all diagonal terms
equal to $1$, and all off-diagonal terms $\tau=0.1$. Other parameters
in the model are $V_{L}=5.0=V_{R},\, A_{L}=4.0=A_{R},\,\Gamma_{11}=0.5,\,\Gamma_{55}=0.5$,
and the inverse temperature $\beta=10$. The resulting current in
each lead is plotted in Fig. \ref{fig:1}(b) (thick lines), along
with the suitably rescaled bias applied to the leads (dotted lines).
In addition, we plot the steady-state value of the current at each
time, using the LB formula. This serves as a comparison between the
best predictions of the steady-state formalism and our fully time-dependent
method, from which it is instantly apparent that the TD current amplitude
is an order of magnitude greater than that of the LB current. There
is a much sharper peak in the transient current in lead $L$ than
in $R$, because the bias applied in $L$ has the value $V_{L}+A_{L}$
at the switch-on time $t=0$, whereas the corresponding bias in lead
$R$ has the value $V_{R}$. In the long-time regime, the current
signal in each lead has the same form but the signal in $R$ is phase-shifted
by $-\pi/2$ with respect to the signal in $L$. Additionally, `ringing'
oscillations in the current occurring at higher frequencies than the
driving frequency $\Omega_{\alpha}$ can be seen, as also observed
in \citep{JAUHO1994} and \citep{Ridley2015}. We identify the physical
source of these frequencies in the energy gap $\left|V_{\alpha}+\mu-E\right|$
between onsite energies and the constant bias shift in the leads. 

One can also see from Fig. \ref{fig:1}(b) that, whereas the LB formalism
preserves the condition of charge conservation, $I_{L}\left(t\right)+I_{R}\left(t\right)=0$,
at all times, this condition is not satisfied by the two TD currents.
We understand this as follows: electrons propagate in the lead at
the finite Fermi velocity $v_{\alpha}^{F}$, so that the effects of
a bias applied locally will take a finite time to propagate to other
parts of the junction \citep{RVLS2013}. If a constant bias is switched
on in each lead, then following a time delay equal to the transmission
time of this signal across the junction, the rate of flow of electrons
from $C$ to $L$ ($I_{L}$) must equal the rate of electron transport
from $R$ to $C$ (-$I_{R}$), as in this case the number of electrons
in the molecular region $N_{C}\left(t\right)=-i\mbox{Tr}\left[\mathbf{G}_{CC}^{<}\left(t,t\right)\right]$
is constant in time. If the bias is time-dependent, then the Hamiltonian
is no longer symmetric under time translations and, in general, $N_{C}\left(t\right)$
also varies with time. Therefore local charge conservation does not
apply in this case, a fact that is reflected in the current characteristics
of Fig. \ref{fig:1}(b). 

\begin{figure}
\includegraphics[scale=0.4]{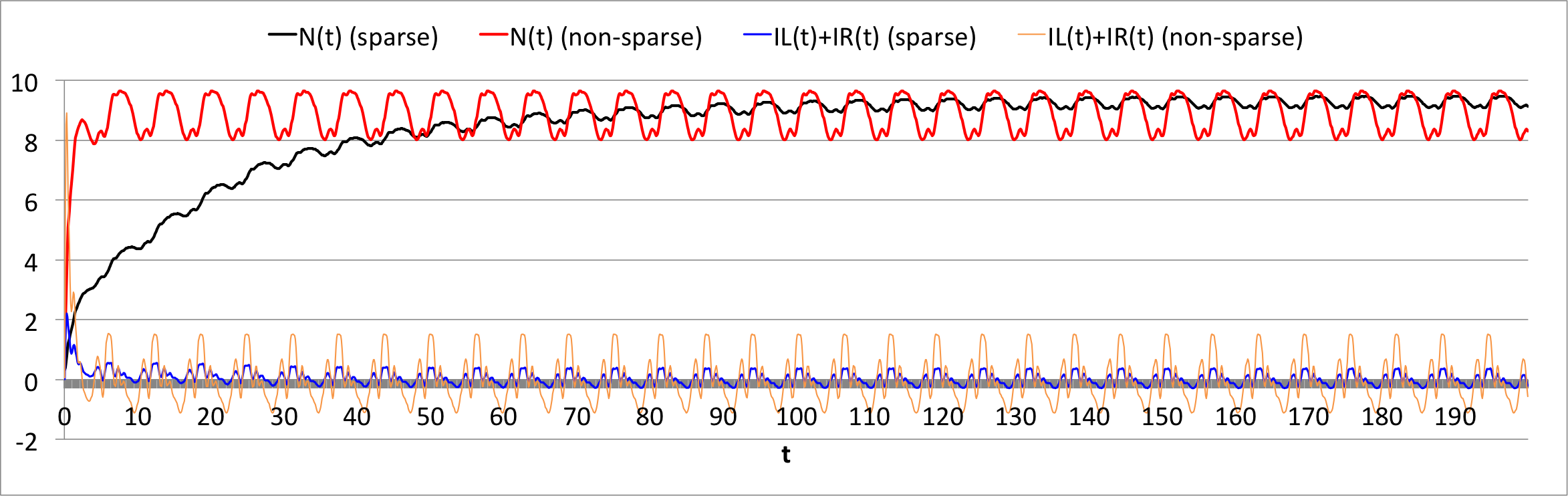}\caption{The electron number (black line) and sum of currents (blue line) for
the sparse tight-binding hamiltonian is compared with the electron
number (red line) and sum of currents (orange line) for a hamiltonian
in which each lead is coupled with the same strength to every site
in the chain. \label{fig:2}}
\end{figure}

It is a point of concern with Fig. \ref{fig:1}(b) that it appears
to indicate an unbroken net flow of charge into the molecule, because
after the usual transient `ringing' regime is over, the average value
of $I_{L}\left(t\right)+I_{R}\left(t\right)$ \textit{per cycle} appears
to be positive. This is a property which, if sustained indefinitely,
means that there will always be a charge increase in the central region
of the junction - clearly an unphysical situation. To investigate
what is occurring here, we compute in Fig. \ref{fig:2} the transport
properties over a much longer time range of the same 5-site model
as was considered in Fig. \ref{fig:1}(b). The long time plots of
both the sum of the currents (blue curve) and the particle number
$N_{C}$ (black) are displayed in Fig. \ref{fig:2}, for this system,
where to compute $N_{C}$ we have expanded (\ref{eq:numbermol}) into
the left/right eigenbasis in a similar fashion to (\ref{eq:CURRENTFINAL}).
Certain qualitative features of the plots stand out, in particular
a periodic `sloshing' of the charge in the molecular region as the
number of particles increases there, before levelling off at a value
below the maximum charge of 10. Simultaneously with this dynamic filling
of the molecular levels, the average value of $I_{L}\left(t\right)+I_{R}\left(t\right)$
drops until the integral under this curve evaluates to zero over a
full cycle. Thus in addition to the fast decay of the initial transient,
there is a secondary underlying decay which occurs over a much longer
time scale. Note that this decay over two time scales is not possible
with a single-level model, so that it is fundamentally a finite-size
effect.

In Fig. \ref{fig:2} we compare the results obtained for the sparse
tight-binding level-width matrix with the same plots for level-widths
that are relatively non-sparse. Specifically, we take $\left[\mathbf{\Gamma}_{L}\right]_{ij}=\delta_{ij}0.5=\left[\mathbf{\Gamma}_{R}\right]_{ij}$,
giving a level-width with a constant term along the main diagonal.
The timescale for decay of all transient modes in formula (\ref{eq:CURRENTFINAL})
is $\tau\equiv Max\{1/\gamma_{j}\}$, where the $\gamma_{j}$ are
defined as imaginary parts of the complex eigenvalues, $\bar{\varepsilon}_{j}=\lambda_{j}-i\gamma_{j}$.
\textcolor{black}{For the sparse tight-binding level-width $\tau=61.9$,
whereas in the non-sparse case we have a decay time of $\tau=2.0$.
}This order-of-magnitude difference is well illustrated by the extremely
fast relaxation of the non-sparse number density (red curve) to a
steady oscillation below $N_{C}=10$. This implies that the integral
of $I_{L}\left(t\right)+I_{R}\left(t\right)$ over one cycle is zero,
implying no long-term change in the sum of currents, as shown by the
orange curve in Fig. (\ref{fig:2}). Additionally, we note that despite
the qualitatively similar `sloshing' between two values of the electron
number in the molecule after $t\sim\tau$ has elapsed, there is a
difference in the amplitude of the oscillating signal for the sparse
and non-sparse cases. This is simply due to the fact that the overlap
integral prefactors (for example $\left\langle \varphi_{j}^{L}\right|\mathbf{\Gamma}_{\alpha}\left|\varphi_{j}^{R}\right\rangle $)
are smaller on average when $\mathbf{\Gamma}_{\alpha}$ is sparse.

\section{Conclusions}

In this work we have shown that the general formula for the current
in a multi-terminal junction derived in \citep{Ridley2015} can be
manipulated into another form which is very convenient for numerical
calculations. Using the Pad\'{e} approximation for the Fermi function
it is possible to convert infinite frequency integrals into fast-converging
summations. In the case of a sinusoidal bias applied to the leads
all the time integrals can be removed analytically and expressed in
terms of special functions. To illustrate this formalism, we considered
a molecular chain in a two-terminal junction with a $-\pi/2$ relative
phase shift in the sinusoidal biases applied to both leads. The calculated
current demonstrates similar transient and ringing effects to those
observed previously \citep{JAUHO1994,Ridley2015} and reveals novel
effects due to this symmetry-breaking. Specifically, we demonstrate
the violation of local charge conservation in the molecular region.
We anticipate that in future we will be able to apply this formalism
to the study of quantum `pumping', where symmetry-breaking parameters
in the periodic driving signal lead to a net transfer of charge through
the system \citep{drivennoise2}. We also find that for extended systems
there can be two distinct decay processes in the current characteristics:
an initial `ringing' transient following the bias switch-on, and also
an underlying decay of the entire signal implied by global charge
conservation. The first type of decay is a well-known effect, and
has been studied before for static \citep{RVL2014} and TD \citep{Ridley2015}
biases. The latter type of decay is characterized by the fact that
the signal retains its shape throughout, and also by a decay time
which is related to sparseness of the level-width matrix. We anticipate
that this double decay effect will be observable in more realistic
extended systems than the one considered here, and that the formalism
developed here provides a fast way to access transient properties
in those systems.

\paragraph{Acknowledgements}

Michael Ridley was supported through a studentship in the Centre for
Doctoral Training on Theory and Simulation of Materials at Imperial
College funded by the Engineering and Physical Sciences Research Council
under grant number EP/G036888/1.


\begin{thebibliography}{10}

\bibitem{LANDAUER1970}
R.~Landauer.
\newblock {\em Phil. Mag.}, 21(172):863, 1970.

\bibitem{BUTTIKER1992}
M.~B\"uttiker.
\newblock {\em Phys. Rev. B}, 46:12485, 1992.

\bibitem{diventra}
M.~Di~Ventra.
\newblock {\em Electrical transport in nanoscale systems}.
\newblock Cambridge University Press, 2008.

\bibitem{RVLS2013}
G.~Stefanucci and R.~van Leeuwen.
\newblock {\em Nonequilibrium Many-Body Theory of Quantum Systems}.
\newblock CUP, 2013.

\bibitem{JAUHO1994}
A.-P. Jauho, N.~S. Wingreen, and Y.~Meir.
\newblock {\em Phys. Rev. B}, 50:5528, 1994.

\bibitem{RVL2013}
R.~Tuovinen, R.~van Leeuwen, E.~Perfetto, and G.~Stefanucci.
\newblock {\em J. Phys.: Conf. Ser.}, 427:012014, 2013.

\bibitem{RVL2014}
R.~Tuovinen, E.~Perfetto, G.~Stefanucci, and R.~van Leeuwen.
\newblock {\em Phys. Rev. B}, 89:085131, 2014.

\bibitem{Ridley2015}
M.~Ridley, A.~MacKinnon, and L.~Kantorovich.
\newblock {\em Phys. Rev. B}, 91:125433, Mar 2015.

\bibitem{StefALm2004}
G.~Stefanucci and C.-O. Almbladh.
\newblock {\em Phys. Rev. B}, 69:195318, 2004.

\bibitem{konstantinov-perel-JETP-1961}
O.~V. Konstantinov and V.~I. Perel'.
\newblock {\em Sov. Phys. JETP}, 12(1):142, 1961.

\bibitem{Li2004}
S.~D. Li, Z.~Yu, S.-F. Yen, W.~C. Tang, and P.~J. Burke.
\newblock {\em Nano Lett.}, 4(4):753, 2004.

\bibitem{RVLS2014}
S.~Latini, E.~Perfetto, A.-M. Uimonen, R.~van Leeuwen, and G.~Stefanucci.
\newblock {\em Phys. Rev. B}, 89:075306, 2014.

\bibitem{lerch1887note}
M.~Lerch.
\newblock {\em Acta Math.}, 11(1-4):19--24, 1887.

\bibitem{Ozaki2007}
T.~Ozaki.
\newblock {\em Phys. Rev. B}, 75:035123, Jan 2007.

\bibitem{Hu2010}
J.~Hu, R.~Xu, and Y.~Yan.
\newblock {\em J. Chem. Phys.}, 133:101106, 2010.

\bibitem{Hu2011}
J.~Hu, M.~Luo, F.~Jiang, R.~Xu, and Y.~Yan.
\newblock {\em J. Chem. Phys.}, 134:244106, 2011.

\bibitem{zhang(b)2013}
Y.~Zhang, S.~Chen, and G.~Chen.
\newblock {\em Phys. Rev. B}, 87(8):085110, 2013.

\bibitem{MUJICA(a)1994}
V.~Mujica, M.~Kemp, and M.~A. Ratner.
\newblock {\em J. Chem. Phys.}, 101:6849, 1994.

\bibitem{MUJICA(b)1994}
V.~Mujica, M.~Kemp, and M.~A. Ratner.
\newblock {\em J. Chem. Phys.}, 101:6856, 1994.

\bibitem{chen2013QDARRAY}
S.~Chen, H.~Xie, Y.~Zhang, X.~Cui, and G.~Chen.
\newblock {\em Nanoscale}, 5(1):169, 2013.

\bibitem{drivennoise2}
M.~Strass, P.~H{\"a}nggi, and S.~Kohler.
\newblock {\em Phys. Rev. Lett.}, 95(13):130601, 2005.

\end{thebibliography}
\end{document}